\documentclass[12pt,a4paper]{article}
\usepackage{t1enc} \usepackage{times}
\usepackage{amssymb}
\usepackage{amsmath}
\usepackage[mathscr]{euscript}
\usepackage{hyperref}
\usepackage[totalwidth=437pt, totalheight=664pt]{geometry}
\DeclareMathAlphabet\scr{U}{scr}{m}{n}
\SetMathAlphabet\scr{bold}{U}{scr}{b}{n}
  \DeclareFontFamily{U}{scr}{\skewchar\font'177}%
  \DeclareFontShape{U}{scr}{m}{n}{<-6>rsfs5<6-8>rsfs7<8->rsfs10}{}%
  \DeclareFontShape{U}{scr}{b}{n}{<-6>rsfs5<6-8>rsfs7<8->rsfs10}{}%

\newcommand{\rr}{\mathbb R}  
\newcommand{\rp}{\mathbb R _+}

\newtheorem{satz}{Theorem}[section]

\newtheorem{cor}[satz]{Corollary}
\newtheorem{@definition}[satz]{Definition}
\newenvironment{defi}{\begin{@definition}\rm}{\end{@definition}}
\newtheorem{@bsp}[satz]{Example}
\newenvironment{bsp}{\begin{@bsp}\rm}{\end{@bsp}}
\newtheorem{@assumption}[satz]{Assumption}

\newtheorem{@convention}[satz]{Convention}

\newtheorem{@remark}[satz]{Remark}
\newenvironment{bem}{\begin{@remark}\rm}{\end{@remark}}

\newcommand{\be}{\begin{enumerate}}
\newcommand{\ee}{\end{enumerate}}
\newcommand{\beq}{\begin{equation}}
\newcommand{\eeq}{\end{equation}}
\newcommand{\bea}{\begin{eqnarray}}
\newcommand{\eea}{\end{eqnarray}}
\newcommand{\beaa}{\begin{eqnarray*}}
\newcommand{\eeaa}{\end{eqnarray*}}

\newcommand{\bpf}{\noindent {\sc Proof.}\ }
\newcommand{\ep}{\hfill $\square $}

\newcommand{\til}{\widetilde}

\renewcommand{\emptyset}{\varnothing}

\renewcommand{\epsilon}{\varepsilon}
\renewcommand{\theta}{\vartheta}
\renewcommand{\rho}{\varrho}

\renewenvironment{thebibliography}[1]{%
\begin{oldthebibliography}{#1}%
\setlength{\baselineskip}{.9em}
\linespread{.9}
\small
\setlength{\parskip}{0ex}%
\setlength{\itemsep}{.1em}%
}%
{%
\end{oldthebibliography}%
}

\begin{document}
\title{Existence of Shadow Prices in\\ Finite Probability Spaces}
\author{Jan Kallsen\footnote{Mathematisches Seminar,
Christian-Albrechts-Universit\"at zu Kiel,
Westring 383,
D-24118 Kiel, Germany,
(e-mail: kallsen@math.uni-kiel.de).}  
\quad Johannes Muhle-Karbe\footnote{Departement Mathematik,
ETH Z\"urich,
R\"amistrasse 101,
CH-8092 Z\"urich, Switzerland,
(e-mail: johannes.muhle-karbe@math.ethz.ch).} 
}
\date{}
\maketitle

\begin{abstract}
A \textit{shadow price} is a process $\til{S}$ lying within the bid/ask prices $\underline{S},\overline{S}$ of a market with proportional transaction costs, such that maximizing expected utility from consumption in the frictionless market with price process $\til{S}$ leads to the same maximal utility as in the original market with transaction costs. For finite probability spaces, this note provides an elementary proof for the existence of such a shadow price.\\

Key words: transactions costs, portfolio optimization, shadow price\\

Mathematics Subject Classification (2000): 91B28, 91B16
\end{abstract}

\section{Introduction}
\markboth{Chapter \ref{c:exshadow}. On the existence of shadow prices in finite discrete time}{\ref{s:exintro}. Introduction}
When considering problems in Mathematical Finance, one classically works with a \textit{frictionless} market, i.e., one assumes that securities can be purchased and sold for the same price $S$. This is clearly a strong modeling assumption, since in reality one usually has to pay a higher \textit{ask price} when purchasing securities, whereas one only receives a lower \textit{bid price} when selling them. Put differently, one is faced with \emph{proportional transaction costs}. The introduction of even miniscule transaction costs often fundamentally changes the structure of the problem at hand (cf., e.g., \cite{davis.norman.90,guasoni.al.07b,cvitanic.al.99c}). Therefore models with transaction costs have been extensively studied in the literature (see, e.g., the recent monograph \cite{kabanov.safarian.09} and the references therein).  

Optimization problems involving transaction costs are usually tackled by one of two different approaches. Whereas the first method employs methods from stochastic control theory, the second reformulates the task at hand as a similar problem in a frictionless market. This second approach goes back to the pioneering paper of Jouini and Kallal \cite{jouini.kallal.95}. They showed that under suitable conditions, a market with bid/ask prices $\underline{S},\overline{S}$ is arbitrage free if and only if there exists a \textit{shadow price} $\til{S}$ lying within the bid/ask bounds, such that the frictionless market with price process $\til{S}$ is arbitrage free.  The same idea has since been employed extensively leading to various other versions of the fundamental theorem of asset pricing in the presence of transaction costs (cf., e.g., \cite{schachermayer.04a,guasoni.al.07b} and the references therein). It has also found its way into other branches of Mathematical Finance. For example, \cite{lamberton.al.98} have shown that bid/ask prices can be replaced by a shadow price in the context of local risk-minimization, whereas \cite{cvitanic.karatzas.96b,cvitanic.wang.01,loewenstein.00,kallsen.muhlekarbe.08c} prove that the same is true for portfolio optimization in certain It\^o process settings. In these articles the duality theory for frictionless markets is typically applied to a shadow price, i.e., shadow prices and the corresponding martingale measures -- \emph{consistent price systems} in the terminology of \cite{schachermayer.04a, guasoni.al.07b} -- play the role of martingale measures in frictionless markets in markets with proportional transaction costs.

In the present study we establish that \emph{in finite probability spaces}, this general principle indeed holds true literally for investment/consumption problems, i.e., \emph{a shadow price always exists}. We first introduce our finite market model with proportional transaction costs in Section \ref{s:dsetup}. Subsequently, we state our main result concerning the existence of shadow prices and prove it using elementary convex analysis.

For a vector $x=(x^1,\ldots,x^d)$, we write $x^+=(\max\{x^1,0\},\ldots,\max\{x^d,0\})$ and $x^-=(\max\{-x^1,0\},\ldots,\max\{-x^d,0\})$. Likewise, inequalities and equalities are understood to be componentwise  in a vector-valued context. Moreover, for any stochastic process $X$ we write $\Delta X_t := X_t-X_{t-1}$.

\section{Utility maximization with transaction costs in finite discrete time}\label{s:dsetup}\index{discrete-time model}\index{proportional transaction costs}
\markboth{Chapter \ref{c:exshadow}. On the existence of shadow prices in finite discrete time}{\ref{s:dsetup}. Utility maximization with transaction costs in finite discrete time}
\setcounter{equation}{0}
We study the problem of maximizing expected utility from consumption in a finite market model with proportional transaction costs.
Our general framework is as follows. Let $(\Omega,\scr{F},(\scr{F}_t)_{t \in \{0,1,\dots,T\}},P)$ be a filtered probability space, where $\Omega=\{\omega_1,\dots,\omega_K\}$ and the time set $\{0,1,\dots,T\}$ are finite. In order to avoid lengthy notation, we let $\scr{F}=\scr{F}_T=\scr{P}(\Omega)$, $\scr{F}_0=\{\emptyset,\Omega\}$, and assume that $P(\{\omega_k\})>0$ for all $k \in \{1,\dots,K\}$. However, one can show that all following statements remain true without these restrictions. 

The financial market we consider consists of a risk-free asset $0$ (also called \textit{bank account}) with price process $S^0$ normalized to $S^0_t=1$, $t=0,\ldots, T$, and risky assets  $1,\dots,d$ whose prices are expressed in multiples of $S^0$. More specifically, they are modelled by their (discounted) \textit{bid price process} $\underline{S}=(\underline{S}^1, \ldots, \underline{S}^d)$ and their (discounted) \textit{ask price process} \index{bid-price} \index{ask-price} $\overline{S}=(\overline{S}^1,\ldots,\overline{S}^d)$, where we naturally assume that  $\overline{S},\underline{S}$ are adapted and satisfy $\overline{S} \geq \underline{S}>0$. Their meaning should be obvious: if one wants to purchase security $i$ at time $t$, one must pay the higher price $\overline{S}^i_t$ whereas one receives only $\underline{S}^i_t$ for selling it. 

The connection to \emph{proportional transaction costs} is the following. In frictionless markets, one models the (mid) price process $S$ of the assets under consideration. Transaction costs equal to a fraction $\overline{\epsilon} \in [0,\infty), \underline{\epsilon} \in [0,1)$ of the amount transacted for purchases and sales of stocks, respectively, then lead to an ask price of $\overline{S}:=(1+\overline{\epsilon})S$ and a bid price of $\underline{S}:=(1-\underline{\epsilon})S$. However, the mid price $S$ does not matter for the modelling of the market with transaction costs, since shares are only bought and sold at $\overline{S}$ resp.\ $\underline{S}$. Therefore we work directly with the bid and ask price processes.

\begin{bem}
 Our setup amounts to assuming that the risk-free asset can be purchased and sold without incurring any transaction costs. This assumption is commonly made in the literature dealing with optimal portfolios in the presence of transaction costs (cf., e.g., \cite{davis.norman.90}), and seems reasonable when thinking of security $0$ as a bank account. For foreign exchange markets where it appears less plausible, a numeraire free approach has been introduced by \cite{kabanov.99}. This approach would, however, require the use of multidimensional utility functions as in \cite{deelstra.al.01,campi.owen.10} in our context. 
\end{bem}

\begin{defi}\index{trading strategy}
A \textit{trading strategy} is an $\rr^{d+1}$-valued predictable stochastic process $(\varphi^0,\varphi)=(\varphi^0,(\varphi^1,\ldots,\varphi^d))$, where $\varphi^i_{t}$, $i=0,\ldots,d$, $t=0,\ldots,T+1$ denotes the number of shares held in security $i$ until time $t$ after rearranging the portfolio at time $t-1$. A (discounted) \emph{consumption process}\index{consumption process} is an $\rr$-valued, adapted stochastic process $c$, where $c_t$, $t=0,\ldots,T$ represents the amount consumed at time $t$. A pair $((\varphi^0,\varphi),c)$ of a trading strategy $(\varphi^0,\varphi)$ and a consumption process $c$ is called \emph{portfolio/consumption pair}\index{portfolio/consumption pair}.
\end{defi}

To capture the notion of a self-financing strategy, we use the intuition that no funds are added or withdrawn. More specifically, this means that the proceeds of selling stock must be added to the bank account while the expenses from consumption and the purchase of stock have to be deducted from the bank account whenever the portfolio is readjusted from $\varphi_t$ to $\varphi_{t+1}$ and an amount $c_t$ is consumed at time $t \in \{0,\ldots,T\}$. Defining purchase and sales processes $\Delta \varphi^{\uparrow}, \Delta \varphi^{\downarrow}$ as
\begin{equation}\label{e:purchases}
\Delta \varphi^{\uparrow}:=(\Delta \varphi)^+, \quad \Delta \varphi^{\downarrow}:=(\Delta \varphi)^-,
\end{equation}
this leads to the following notion.

\begin{defi}
A portfolio/consumption pair $((\varphi^0,\varphi),c)$ is called \emph{self-financing}\index{portfolio/consumption pair!self-financing} (or $(\varphi^0,\varphi)$ {\em $c$-financing})\index{trading strategy!$c$-financing} if
\begin{equation}\label{e:dselff1}
\Delta\varphi^0_{t+1}=\underline S^{\top}_t\Delta\varphi^{\downarrow}_{t+1} - \overline S^{\top}_t\Delta\varphi^{\uparrow}_{t+1}-c_t, \quad t=0,\ldots,T. 
\end{equation}
\end{defi}

\begin{bem} \label{bem:dselff2}
Define the cumulated purchases $\varphi^{\uparrow}$ and sales $\varphi^{\downarrow}$ as
\begin{align*}
\varphi^{\uparrow}_t			:= (\varphi_0)^+ +\sum_{s=1}^t \Delta \varphi^{\uparrow}_s, \quad 
\varphi^{\downarrow}	_t	:= (\varphi_0)^- +\sum_{s=1}^t \Delta \varphi^{\downarrow}_s, \quad t=1,\ldots,T+1.
\end{align*}
Then the self-financing condition \eqref{e:dselff1}  implies that $\left((\varphi^0,\varphi^{\uparrow},- \varphi^{\downarrow}),c\right)$ is self-financing in the usual sense for a frictionless  market with $2d+1$ securities $(1,\overline S,\underline S)$. Moreover, note that for $\underline S=\overline S$, we recover the usual self-financing condition for frictionless markets.
\end{bem}

We consider an investor who disposes of an \textit{initial endowment} $(\eta_0, \eta)\in \rp^{d+1}$, referring to the initial number of securities of type $i$, $i=0,\ldots,d$, respectively.  

\begin{defi}\label{d:doptimal}\index{portfolio/consumption pair!admissible}
A self-financing portfolio/consump\-tion pair $((\varphi^0,\varphi),c)$ is called \emph{admissible} if  $(\varphi^0_0,\varphi_0)=(\eta_0,\eta)$ and $(\varphi^0_{T+1},\varphi_{T+1})=(0,0)$. An admissible portfolio/consumption pair $((\varphi^0,\varphi),c)$ is called \emph{optimal} if it maximizes
\begin{equation}\label{e:doptimal}
\kappa \mapsto E\left(\sum_{t=0}^T u_t(\kappa_t)\right)
\end{equation}
over all admissible portfolio/consumption pairs $((\psi^0,\psi),\kappa)$, where the \textit{utility process} $u$ is a mapping $u: \Omega \times \{0,\ldots,T\} \times \rr \to [-\infty,\infty)$, such that $(\omega,t) \mapsto u_t(\omega,x)$ is predictable for any $x \in \rr$ and $x \mapsto u_t(\omega,x)$ is a proper (in the sense of Rockafellar \cite{rockafellar.97}), upper-semicontinuous, concave function for any $(\omega,t) \in \Omega \times \{0,\ldots,T\}$, which is increasing on its convex effective domain $\{x \in \rr: u_t(\omega,x)>-\infty\}$ for $(\omega,t)$, $t \in \{0,\ldots,T-1\}$ and strictly increasing for $(\omega,T)$ . 
\end{defi}

In view of Definition \ref{d:doptimal}, we only deal with portfolio/consumption pairs where the \emph{entire} liquidation wealth of the portfolio is consumed at time $T$. Note that this can be done without loss of generality,  because the utility process is increasing in consumption.

\begin{bem}
Since we allow the utility process to be random, assuming $S^0_t=1$, $t=0,\ldots,T$ also does not entail a loss of generality in the present setup. 
More specifically, let $S^0$ be an arbitrary strictly positive, predictable process. In this undiscounted case a portfolio/consumption pair $(\varphi,c)$ should be called \textit{self-financing} if
\begin{equation*}
\Delta\varphi^0_{t+1}S^0_t=\underline S^{\top}_t\Delta\varphi^{\downarrow}_{t+1} - \overline S^{\top}_t\Delta\varphi^{\uparrow}_{t+1}-c_t, 
\end{equation*}
for $t=0,\ldots,T$. \textit{Admissibility} is defined as before. By direct calculations, one easily verifies that $((\varphi^0,\varphi),c)$ is self-financing resp.\ admissible if and only if $((\varphi^0,\varphi),\hat{c})=((\varphi^0,\varphi),c/S^0)$ is self-financing resp.\ admissible relative to the discounted processes $\hat{S}^0:=S^0/S^0=1$, $\hat{\overline{S}}:=\overline{S}/S^0$ and $\hat{\underline{S}}:=\underline{S}/S^0$. In view of 
\begin{equation*}
E\left(\sum_{t=0}^T u_t(c_t)\right)=E\left(\sum_{t=0}^T \hat{u}_t(\hat{c}_t)\right)
\end{equation*}
for the utility process $\hat{u}_t(x)=u_t(S^0 x)$, the problem of maximizing undiscounted utility with respect to $u$ is equivalent to maximizing discounted expected utility with respect to $\hat{u}$. 
\end{bem}

We now mention some well-known specifications that are included in our setup.

\begin{bsp}
\begin{enumerate}
\item Maximizing expected \emph{utility from terminal wealth} at time $T$ is included as a special case by setting 
$$u_t(x)=\begin{cases} -\infty, &\mbox{for } x <0,\\ 0, &\mbox{for } x \geq 0, \end{cases} \quad \mbox{for } t \in \{0,\ldots,T-1\}.$$
\item One also obtains a utility process in the sense of Definition \ref{d:doptimal} via 
$$u(\omega,t,x):=D_t(\omega) u(x),$$
where $D$ is some positive predictable discount factor (e.g., $D_t=\exp(-r t)$ or $D_t=1/(1+r)^t$ for $r>0$) and $u:\rr\to \rr \cup \{-\infty\}$ is a utility function in the usual sense, as, e.g., the logarithmic utility function $u(x)=\log(x)$, a power utility function $u(x)=x^{1-p}/(1-p)$, $p \in \mathbb{R}_+ \backslash \{0,1\}$, or an exponential utility function $u(x)=e^{-px}/p$, $p>0$.
\end{enumerate}
\end{bsp}

In particular, one does not have to rule out negative consumption from a mathematical point of view, even though allowing it seems rather dubious from an economical perspective.

\section{Existence of shadow prices}\label{s:dmainresult}
\markboth{Chapter \ref{c:exshadow}. On the existence of shadow prices in finite discrete time}{\ref{s:dmainresult}. Existence of shadow prices}
\setcounter{equation}{0}

We now introduce the central concept of this paper.

\begin{defi}\index{shadow price}
We call an adapted process $\til S$ \emph{shadow price process} if 
$$\underline S \leq \til{S} \leq \overline S$$
and if the maximal expected utilities in the market with bid/ask-prices $\underline S, \overline S$ and in the market with price process $\til S$ {\em without} transaction costs coincide.
\end{defi}

The following theorem shows that in our finite market model, shadow price processes always exist, except in the trivial case where all admissible portfolio/consumption pairs lead to expected utility $-\infty$. The main idea of the proof is to treat purchases and sales separately in \eqref{e:dselff1}. This means that we effectively consider a problem with two sets of assets whose holdings must be in- resp.\ decreasing. Maybe surprisingly, the Lagrange multipliers corresponding to these constraints merge into only one process (rather than two). The latter has a natural interpretation
as a shadow price process.

\begin{satz}\label{dshadow}
Suppose an optimal portfolio/consumption pair $((\varphi^0,\varphi),c)$ exists for the market with bid/ask prices $\underline{S},\overline{S}$. Then if $E(\sum_{t=0}^T u_t(c_t))>-\infty$, a shadow price process $\til{S}$ exists.  
\end{satz}

\bpf 
\emph{Step 1}: As the utility process is increasing, allowing for sales and purchases at the same time does not increase the maximal expected utility. More precisely, since $x \mapsto u_t(x)$ is increasing for fixed $t$, maximizing \eqref{e:doptimal} over all admissible portfolio/consumption pairs yields the same maximal expected utility as maximizing \eqref{e:doptimal} over the set of all $((\psi^0,\psi^{\uparrow},\psi^{\downarrow}),\kappa)$, where $(\psi^0(t))_{t=0,\ldots,T+1}$ is an $\rr$-valued predictable process with $\psi^0_0=\eta_0$ and $\psi^0_{T+1}=0$, the increasing, $\rr^d$-valued predictable processes $(\psi^{\uparrow}_t)_{t=0,\ldots,T+1}$, $(\psi^{\downarrow}_t)_{t=0,\ldots,T+1}$ satisfy $\psi^{\uparrow}_0=\eta^+$, $\psi^{\downarrow}_0=\eta^-$, $\psi^{\uparrow}_{T+1}-\psi^{\downarrow}_{T+1}=0$ and $(\kappa_t)_{t=0,\ldots,T}$ is a consumption process such that \eqref{e:dselff1} holds for $t=0,\ldots,T$ and $((\psi^0,\psi),\kappa)$ instead of $((\varphi^0,\varphi),c)$. Moreover, if we define $\Delta \varphi^{\uparrow}$ and $\Delta \varphi^{\downarrow}$ as in \eqref{e:purchases} above and set
\begin{align*}
\varphi^{\uparrow}  :=\eta^++\sum_{t=1}^\cdot \Delta\varphi^{\uparrow}_t, \quad \varphi^{\downarrow}:=\eta^-+\sum_{t=1}^\cdot \Delta \varphi^{\downarrow}_t,
\end{align*}
then $((\varphi^0,\varphi^{\uparrow},\varphi^{\downarrow}),c)$ is an optimal strategy in this set.

\textit{Step 2}: We now formulate our optimization problem as a finite-dimensional convex minimization problem with convex constraints. To this end, denote by $F^1_{t},\ldots, F^{m_t}_{t}$ the partition of $\Omega$ that generates $\scr{F}_t$, $t \in \{0,\ldots,T\}$. Since a mapping is $\scr{F}_t$-measurable if and only if it is constant on the sets $F^j_{t}$, $j=1,\ldots,m_t$, we can identify the set of all processes $((\psi^0,\psi^{\uparrow},\psi^{\downarrow}),\kappa)$, where $(\psi^0_t)_{t=0,\ldots,T+1}$ is $\rr$-valued and predictable with $\psi^0_0=\eta_0$, $(\psi^{\uparrow}_t)_{t=0,\ldots,T+1}$ and $(\psi^{\downarrow}_t)_{t=0,\ldots,T+1}$ are increasing, $\rr^d$-valued and predictable with $\psi_0^{\uparrow}=\eta^+$, $\psi_0^{\downarrow}=\eta^-$ and $(\kappa_t)_{t=0,\ldots,T}$ is a consumption process such that \eqref{e:dselff1} holds for $t=0,\ldots,T$ with  
\begin{equation*}
\rp^{2dn } \times \rr^n:=(\rp^{m_0 d} \times \ldots \times \rp^{m_T d})\times(\rp^{m_0 d} \times \ldots \times \rp^{m_T d})\times (\rr^{m_0} \times \ldots \times \rr^{m_T}),
\end{equation*}
and vice versa, namely with 
\begin{equation*}
\begin{split}
(\Delta \psi^{\uparrow},\Delta \psi^{\downarrow},c):=(\Delta\psi^{\uparrow,1,1}_1,\ldots,\Delta \psi^{\uparrow,m_T,d}_{T+1},\Delta\psi^{\downarrow,1,1}_1,\ldots,\Delta\psi^{\downarrow,m_T,d}_{T+1}, c^1_0,\ldots,c^{m_T}_T),
\end{split}
\end{equation*}
where we use the notation $\Delta\psi^{\uparrow,j,i}_t:= \Delta\psi^{\uparrow,i}_t(\omega)$ for $i=1,\ldots,d$, $t=0,\ldots,T$, $j=1,\ldots,m_t$, and $\omega \in F^j_{t}$ (and analogously for $\Delta\psi^{\downarrow}$, $c$, $\underline{S}$, $\overline{S}$). Using this identification, we can define mappings  $f: \rp^{2dn } \times \rr^n \to \rr \cup \{\infty\}$, $h_0^{j}: \rp^{2dn} \times \rr^n \to \rr$ and $h^{j}:\rp^{2dn}\times \rr^n \to \rr^d$ (for $j=1,\ldots,m_T$) by
\begin{align*}
f(\Delta \psi^{\uparrow},\Delta \psi^{\downarrow},c)    &:=-E\left(\sum_{t=1}^T u_t(c_t)\right),\\
h_0^{j}(\Delta \psi^{\uparrow},\Delta \psi^{\downarrow},c)  &:= \eta_0+\sum_{t=1}^T \left( (\underline{S}^{j}_{t-1})^{\top}\Delta\psi^{\downarrow,j}_t-(\overline{S}^{j}_{t-1})^{\top}\Delta\psi_t^{\uparrow,j}\right)-\sum_{t=0}^T c^j_t,\\
h^{j}(\Delta\psi^{\uparrow},\Delta\psi^{\downarrow},c) &:=\eta+\sum_{t=1}^{T+1} \left(\Delta \psi^{\uparrow,j}_t-\Delta \psi^{\downarrow,j}_t\right).
\end{align*}
Note that $h_0$ resp.\ $h$ represent the terminal positions in bonds resp.\ stocks. With this notion, $(\Delta\varphi^{\uparrow},\Delta\varphi^{\downarrow},c)$ is optimal if and only if it minimizes $f$ over $\rp^{2dn} \times \rr^n$ subject to the constraints $h_0^j=0$ and $h^{j}=0$ for $j=1,\ldots,m_T$. Since all mappings are actually convex functions on $\rr^{(2d+1)n}$, this is equivalent to $(\Delta\varphi^{\uparrow}, \Delta\varphi^{\downarrow},c)$ minimizing $f$ over $\rr^{(2d+1)n}$ subject to the constraints $h_0^j=0$, $h^{j}=0$ (for $j=1,\ldots,m_T$) and $g^{\uparrow,j}_{t}, g^{\downarrow,j}_{t} \leq 0$ (for $t=0,\ldots,T$ and $j=1,\ldots,m_{t}$), where the convex mappings $g^{\uparrow,j}_{t},g^{\downarrow,j}_{t}: \rr^{(2d+1)n} \to \rr^d$ are given by
\begin{equation*}
g_{t}^{\uparrow,j}(\Delta \psi^{\uparrow},\Delta \psi^{\downarrow},c):=-\Delta \psi^{\uparrow,j}_{t+1}, \quad
g_{t}^{\downarrow,j}(\Delta \psi^{\uparrow},\Delta\psi^{\downarrow},c):=-\Delta \psi^{\downarrow,j}_{t+1}.
\end{equation*}
In view of \cite[Theorems 28.2 and 28.3]{rockafellar.97},  $(\Delta\varphi^{\uparrow},\Delta\varphi^{\downarrow},c)$ is therefore optimal if and only if there exists a \emph{Lagrange multiplier}, i.e., real numbers $\nu^j$, $\mu^{j,i}$ (for $i=1,\ldots,d$ and $j=1,\ldots,m_T$) and $\lambda^{\uparrow,j,i}_{t}, \lambda^{\downarrow,j,i}_{t}$ (for $t=0,\ldots,T$, $i=1,\ldots,d$ and $j=1,\ldots,m_{t}$) such that the following holds.
\begin{enumerate}
\item For $t=0,\ldots,T$, $j=1,\ldots,m_{t}$ and $i=1,\ldots,d$, we have $\lambda^{\uparrow,j,i}_{t}, \lambda^{\downarrow,j,i}_{t} \geq 0$ as well as  $g_t^{\uparrow,j,i}(\Delta\varphi^{\uparrow},\Delta \varphi^{\downarrow},c),g_t^{\downarrow,j,i}(\Delta\varphi^{\uparrow},\Delta \varphi^{\downarrow},c)\leq 0$ and  $\lambda^{\uparrow,j,i}_{t} g^{\uparrow,j,i}_{t}(\Delta\varphi^{\uparrow},\Delta \varphi^{\downarrow},c)=0$ as well as  $\lambda^{\downarrow,j,i}_{t} g^{\downarrow,j,i}_{t}(\Delta\varphi^{\uparrow},\Delta\varphi^{\downarrow},c) =0$ .
\item $h_0^j(\Delta\varphi^{\uparrow},\Delta\varphi^{\downarrow},c)=0$ and $h^{j}(\Delta\varphi^{\uparrow},\Delta\varphi^{\downarrow},c)=0$ for $j=1,\ldots,m_T$.
\item 
\begin{align*}
0\in &\partial f(\Delta\varphi^{\uparrow},\Delta\varphi^{\downarrow},c)+ \sum_{j=1}^{m_T} \nu^j \partial h_0^j(\Delta\varphi^{\uparrow},\Delta\varphi^{\downarrow},c)+\sum_{i=1}^d \sum_{j=1}^{m_T} \mu^{j,i} \partial h^{j,i}(\Delta\varphi^{\uparrow},\Delta\varphi^{\downarrow},c)\\
&+\sum_{t=0}^T\sum_{i=1}^d \sum_{j=1}^{m_{t}} \lambda^{\uparrow,j,i}_{t} \partial g^{\uparrow,j,i}_{t}(\Delta\varphi^{\uparrow},\Delta\varphi^{\downarrow},c)+\sum_{t=0}^T\sum_{i=1}^d \sum_{j=1}^{m_{t}} \lambda^{\downarrow,j,i}_{t} \partial g^{\downarrow,j,i}_{t}(\Delta\varphi^{\uparrow},\Delta\varphi^{\downarrow},c).
\end{align*}
\end{enumerate}
Here, $\partial$ denotes the subdifferential of a convex mapping (cf.\ \cite{rockafellar.97} for more details).\index{subdifferential}

\textit{Step 3}: We now use the optimality conditions for the market with transaction costs to construct a shadow price process. By \cite[Proposition 10.5]{rockafellar.wets.98} we can split Statement 3 into many similar statements where the subdifferentials on the right-hand side are replaced with partial subdifferentials relative to $\Delta\varphi^{\uparrow,1,1}_1,\ldots, \Delta \varphi^{\uparrow,m_T,d}_{T+1}$, $\Delta\varphi^{\downarrow,1,1}_1,\ldots, \Delta \varphi^{\downarrow,m_T,d}_{T+1}$, $c^1_t,\ldots,c^{m_T}_T$, respectively. In particular, for $c^j_T$, $j \in \{1,\ldots,m_T\}$, we obtain
\begin{equation}\label{nu}
0 \in \partial_{c^j_T} f(\Delta \varphi^{\uparrow},\Delta \varphi^{\downarrow},c)-\nu^j, 
\end{equation}
where $\partial_x$ denotes the partial subdifferential of a convex function relative to a vector $x$. Hence $\nu^j < 0$, $j=1,\ldots,m_T$, because $f$ is strictly decreasing in $c^j_T$.  Furthermore, since the mappings $g^{\uparrow,j,i}_{t}, g^{\downarrow,j,i}_{t}$ (for $t=0,\ldots, T$,  $j=1,\ldots,m_t$ and $i=1,\ldots, d$) and $h_0^j, h^{j,i}$ (for $j=1,\ldots,m_T$ and  $i =0,\ldots,d$) are differentiable, their partial subdifferentials coincide with the respective partial derivatives by \cite[Theorem 25.1]{rockafellar.97}. Hence, taking partial derivatives with respect to $\Delta\varphi^{\uparrow,j,i}_{t+1}$ resp.\ $\Delta\varphi^{\downarrow,j,i}_{t+1}$, $t \in \{0,\ldots,T\}$, $j \in \{1,\ldots,m_t\}$, $i \in \{0,\ldots,d\}$, Statement 3 above implies that
\begin{equation}\label{e:diff1}
\begin{split}
0&=\sum_{k: \omega_k \in F^j_{t}}\mu^{k,i} -\bigg(\sum_{k: \omega_k \in F^j_{t}}\nu^k\bigg) \overline{S}^{j,i}_t -\lambda^{\uparrow,j,i}_{t}\\
&=\sum_{k: \omega_k \in F^j_{t}}\mu^{k,i} -\left(\sum_{k: \omega_k \in F^j_{t}}\nu^k\right) \left(1+\frac{\lambda^{\uparrow,j,i}_{t}}{\overline{S}^{j,i}_t\sum_{k: \omega_k \in F^j_{t}}\nu^k}\right)\overline{S}^{j,i}_t,
\end{split}
\end{equation}
and likewise
\begin{equation}\label{e:diff2}
\begin{split}
0=\sum_{k: \omega_k \in F^j_{t}}\mu^{k,i} -\left(\sum_{k: \omega_k \in F^j_{t}}\nu^k\right) \left(1-\frac{\lambda^{\downarrow,j,i}_{t}}{\underline{S}^{j,i}_t\sum_{k: \omega_k \in F^j_{t}}\nu^k}\right)\underline{S}^{j,i}_t.
\end{split}
\end{equation}
In particular we have, for $t=0,\ldots,T$, $j=1,\ldots,m_t$, $i=1,\ldots,d$,
\begin{equation*}
\left(1+\frac{\lambda^{\uparrow,j,i}_{t}}{\overline{S}^{j,i}_t\sum_{k: \omega_k \in F^j_{t}}\nu^k}\right)\overline{S}^{j,i}_t= \left(1-\frac{\lambda^{\downarrow,j,i}_{t}}{\underline{S}^{j,i}_t\sum_{k: \omega_k \in F^j_{t}}\nu^k}\right)\underline{S}^{j,i}_t=:\widetilde{S}^{j,i}_t.
\end{equation*}
Since $\til S:=(\til{S}^1,\ldots,\til{S}^d)$ is constant on $F^j_{t}$ by definition, this defines an adapted process.  Furthermore, we have $\underline{S} \leq \til{S} \leq \overline{S}$, since $\lambda^{\uparrow,j,i}_t,\lambda^{\downarrow,j,i}_t \geq 0$, for $i=1,\ldots,d$, $t=0,\ldots,T$ and $j=1,\ldots,m_t$, and because $\nu^k<0$ for $k=1,\ldots,m_T$. Moreover, by Statement 1 above, we have $\lambda^{\uparrow,j,i}_t=0$ if $\Delta \varphi^{\uparrow,j,i}_t>0$ and $\lambda^{\downarrow,j,i}_t=0$ if $\Delta \varphi^{\downarrow,j,i}_t>0$, such that
\begin{equation}\label{trade}
\til{S}^i=\overline{S}^i \mbox{ on } \{\Delta\varphi^{\uparrow,i}>0\}, \quad \til{S}^i=\underline{S}^i \mbox{ on } \{\Delta\varphi^{\downarrow,i}>0\}.
\end{equation}
Set $\til{\mu}^{j,i}:=\mu^{j,i}$ (for $j=1,\ldots,m_T$, $i=1,\ldots,d$), $\til{\nu}^j:=\nu^j$ (for $j=1,\ldots,m_T$) and $\til{\lambda}^{\uparrow,j,i}_{t},\til{\lambda}^{\downarrow,j,i}_{t}:=0$ (for $t=0,\ldots,T$, $j=1,\ldots,m_{t}$ and $i=1,\ldots,d$). Statements $1$, $2$ and $3$ above, Equations \eqref{e:diff1}, \eqref{e:diff2}, \eqref{trade} and the definition of $\til{S}$ then yield the following. 
\begin{enumerate}  
\item  For $t=0,\ldots,T$, $i=1,\ldots,d$ and $j=1,\ldots,m_{t}$ we have $\tilde{\lambda}^{\uparrow,j,i}_{t}, \til{\lambda}^{\downarrow,j,i}_{t} \geq 0$ as well as $\til{g}_t^{\uparrow,j,i}(\Delta\varphi^{\uparrow},\Delta\varphi^{\downarrow},c), \til{g}_t^{\downarrow,j,i}(\Delta\varphi^{\uparrow},\Delta\varphi^{\downarrow},c) \leq 0$  and $\tilde{\lambda}^{\uparrow,j,i}_{t} \tilde{g}^{\uparrow,j,i}_{t}(\Delta\varphi^{\uparrow},\Delta\varphi^{\downarrow},c)=0$ as well as  $\til{\lambda}^{\downarrow,j,i}_{t} \til{g}^{\downarrow,j,i}_{t}(\Delta\varphi^{\uparrow},\Delta\varphi^{\downarrow},c) =0$,
\item  $\til{h}_0^j(\Delta\varphi^{\uparrow},\Delta\varphi^{\downarrow},c)=0$ and $\til{h}^{j}(\Delta\varphi^{\uparrow},\Delta\varphi^{\downarrow},c)=0$ for $j=1,\ldots,m_T$, 
\item 
\begin{align*}
0 \in &\partial \til{f}(\Delta\varphi^{\uparrow},\Delta\varphi^{\downarrow},c)+ \sum_{j=1}^{m_T} \til{\nu}^j \partial \til{h}_0^j(\Delta\varphi^{\uparrow},\Delta\varphi^{\downarrow},c)+\sum_{i=1}^d \sum_{j=1}^{m_T} \til{\mu}^{j,i} \partial \til{h}^{j,i}(\Delta\varphi^{\uparrow},\Delta\varphi^{\downarrow},c) \\
&-\sum_{t=0}^T \sum_{i=1}^d \sum_{j=1}^{m_{t}} \til{\lambda}^{\uparrow,j,i}_{t} \partial \til{g}^{\uparrow,j,i}_{t}(\Delta\varphi^{\uparrow},\Delta\varphi^{\downarrow},c)-\sum_{t=0}^T \sum_{i=1}^d \sum_{j=1}^{m_{t}} \til{\lambda}^{\downarrow,j,i}_{t} \partial \til{g}^{\downarrow,j,i}_{t}(\Delta\varphi^{\uparrow},\Delta\varphi^{\downarrow},c),
\end{align*}
\end{enumerate}
where the mappings $\til{f}$,  $\til{h}_0^j$, $\til{h}^{j}$, $\til{g}^{\uparrow,j}_{t}$, $\til{g}^{\downarrow,j}_{t}$ are defined by setting $\underline{S}=\overline{S}=\til{S}$ in the definition of the mappings $f$, $h_0^j$, $h^{j}$, $g^{\uparrow,j}_{t}$, $g^{\downarrow,j}_{t}$ above. In view of \cite[Theorem 28.3]{rockafellar.97} and Steps 1 and 2 above, $(\varphi,c)$ is therefore not only optimal in the market with bid/ask prices $\underline{S},\overline{S}$, but in the market with bid-ask prices $\til{S},\til{S}$ (i.e., in the frictionless market with price process $\til{S}$) as well. Hence $\til{S}$ is a shadow price process and we are done. \ep


\begin{bem}
Suppose that, for any $\epsilon>0$ and $(\omega,t) \in \Omega \times \{0,\dots,T\}$, there exist $x_1,x_2$ such that $x \mapsto u_t(\omega,x)$ is differentiable at $x_1,x_2$ and $u'_t(\omega,x_1)/u'_t(\omega,x_2)<\epsilon$. Then it follows from standard arguments in convex analysis along the lines of \cite[Lemma 2.9]{kallsen.02} that an optimal portfolio consumption/consumption pair exists if the market does not allow for arbitrage.
\end{bem}

By the fundamental theorem of asset pricing with transaction costs in finite probability spaces (cf.\ \cite{schachermayer.04a}), absence of arbitrage in our model is equivalent to the existence of a \emph{consistent price system}. This is a pair consisting of an adapted process $S$ evolving within the bid-ask spread $[\underline{S},\overline{S}]$ and a corresponding equivalent martingale measure $Q$. Similarly, the following result characterizes the optimal consumption process in terms of a specific consistent price system, namely a shadow price and a specific martingale measure for the corresponding frictionless market. In analogy to the fundamental theorem of asset pricing, we daringly call it a \emph{fundamental theorem of utility maximization with transaction costs}.

\begin{cor}\label{fundamental}
Let $((\varphi^0,\varphi),c)$ be an admissible portfolio consumption pair for the market with bid/ask prices $\underline{S},\overline{S}$ satisfying $E(\sum_{t=0}^T u_t(c_t))>-\infty$. Then we have equivalence between:
\begin{enumerate}
\item $((\varphi^0,\varphi),c)$ is optimal in the market with bid/ask prices $\underline{S},\overline{S}$.
\item There exists a consistent price system $(\til{S},\til{Q})$ and a number $\alpha \in (0,\infty)$ such that
$$E\left(\frac{d\til{Q}}{dP}\bigg|\scr{F}_t\right)\in \frac{1}{\alpha} \partial u_t(c_t), \quad t=0,\ldots,T.$$
\end{enumerate}
\end{cor}

\bpf  $1 \Rightarrow 2$: We use the notation from the proof of Theorem \ref{dshadow}. In particular, $\til{S}$ and $\til{\nu},\til{\mu}$ denote the shadow price and the corresponding Lagrange multipliers introduced there. Since $\til{\nu}^j<0$ for $j=1,\ldots,m_T$, 
$$\til{Q}(F^j_T):=-\til{\nu}^j/\alpha, \quad j=1,\ldots,m_T,$$
with $\alpha:=\sum_{k=1}^{m_T} -\til{\nu}^k$, defines a measure on $\scr{F}$, which is equivalent to $P$ . Moreover, since the Radon-Nikod\'ym density of $\til{Q}$ with respect to $P$ is given by $(d\til{Q}/dP)^j=-\til{\nu}^j/(\alpha P(F^j_T))$, $j=1,\ldots,m_T$, the density process of $\til{Q}$ with respect to $P$ is given by
$$\til{Z}_t^j:=E\left(\frac{d\til{Q}}{dP}\bigg| \scr{F}_t\right)^j=\frac{\sum_{k: \omega_k \in F_t^j} -\nu_k}{\alpha P(F_t^j)}, \quad t=1,\ldots, T, \quad j=1,\ldots,m_t.$$
By considering the partial subdifferentials with respect to $c_t^j$, $t=1,\ldots,T$, $j=1,\ldots,m_t$ in optimality condition 3 for the process $\tilde{S}$ in the proof of Theorem \ref{dshadow}, we find that $\til{Z}_t$ lies in the subdifferential $\frac{1}{\alpha}\partial u_t(c_t)$ for $t=1,\ldots,T$. It therefore remains to show that $\til{Q}$ is a martingale measure for $\til{S}$, i.e., that $\til{Z}\til{S}^i$ is a $P$-martingale for $i=1,\ldots,d$. By definition of $\til{Z}$ resp.\ $\til{S}$ and \eqref{e:diff2}, we have $\til{Z}_T^j\til{S}^{j,i}_T=-\til{\nu}^j \til{S}_T^{j,i}/(\alpha P(F^j_T))=-\mu^{j,i}/(\alpha P(F^j_T))$ for $i=1,\ldots,d$ and $j=1,\ldots,m_T$. Hence $\til{Z}\til{S}$ is a martingale, because, for $i=1,\ldots,d$, $t=0,\ldots,T-1$ and $j=1,\ldots,m_t$, we have
\begin{align*}
E(\til{Z}_T\til{S}_T^i|\scr{F}_t)^j = \frac{\sum_{k: \omega_k \in F_t^j} P(\{\omega_k\})\frac{-\mu^{k,i}}{\alpha P(\{\omega_k\})}}{P(F_t^j)} &=\frac{-(\sum_{k:\omega_k \in F_t^j} \nu_k)\til{S}_t^j}{\alpha P(F_t^j)}=\til{Z}_t^j \til{S}_t^j,
\end{align*}
where we have again used \eqref{e:diff2} for the second equality. 

$2 \Rightarrow 1$: We first show that Statement $2$ implies that $((\varphi^0,\varphi),c)$ is optimal in the frictionless market with price process $\til{S}$. For any admissible portfolio/consumption pair $((\psi^0,\psi),\kappa)$, summing \eqref{e:dselff1} over $t=0,\ldots,T+1$, inserting $(\psi^0_{T+1},\psi_{T+1})=(0,0)$, and using the $\tilde{Q}$-martingale property of $\tilde{S}$ yields the budget constraint
\begin{equation}\label{e:budget}
E_{\til{Q}}\left(\sum_{t=0}^T u_t(\kappa_t)\right) =\eta_0+\eta^\top \til{S}_0.
\end{equation}
In particular, this holds for $((\varphi^0,\varphi),c)$. Now let $((\psi^0,\psi),\kappa)$ be any competing admissible strategy. Since the utility process is concave, we have
\begin{align*}
E\left(\sum_{t=0}^T u_t(\kappa_t)\right) &\leq E\left(\sum_{t=0}^T u_t(c_t)\right)+\alpha E\left(\frac{d\til{Q}}{dP}\left(\sum_{t=0}^T \kappa_t-\sum_{t=0}^T c_t\right)\right),
\end{align*}
by assumption and definition of the subdifferential. Hence \eqref{e:budget} implies that $((\varphi^0,\varphi),c)$ is optimal in the frictionless market with price process $\til{S}$.

Now let $((\psi^0,\psi),\kappa)$ be any admissible portfolio consumption pair in the market with bid/ask prices $\underline{S}, \overline{S}$.  For $t=1,\ldots,T+1$, define $\Delta \psi^{\uparrow}_t:=(\Delta \psi_t)^+$, $\Delta \psi^{\downarrow}_t:=(\Delta \psi_t)^-$ and let
\begin{equation*}
\til{\kappa}(t):=\kappa(t)+ (\Delta \psi^{\uparrow}_t)^{\top}(\overline{S}_t-\til{S}_t)+(\Delta\psi^{\downarrow}_t)^{\top}(\til{S}_t-\underline{S}_t).
\end{equation*}
Then $\til{\kappa}\geq \kappa$ since $\underline{S} \leq \til{S} \leq \overline{S}$ and $((\psi^0,\psi),\til{\kappa})$ is a self-financing portfolio/consumption pair in the frictionless market with price process $\til{S}$, i.e., with bid/ask-prices $\til{S},\til{S}$. Since $((\varphi^0,\varphi),c)$ is optimal in this market, we have
\begin{equation*}
E\left(\sum_{t=0}^T u_t(\kappa_t)\right) \leq E\left(\sum_{t=0}^T u_t(\til{\kappa}_t)\right)\leq E\left(\sum_{t=0}^T u_t(c_t)\right).
\end{equation*}
Therefore $((\varphi^0,\varphi),c)$ is optimal in the market with bid/ask prices $\underline{S},\overline{S}$ as well. \ep \\

\begin{bem} 
If, for fixed $(\omega,t) \in \Omega \times \rp$, the mapping $x \mapsto u_t(\omega,x)$ is differentiable on its effective domain with derivative $u'$, then $E(\frac{d\til{Q}}{dP}|\scr{F}_t) \in \frac{1}{\alpha} \partial u_t(c_t)$ reduces to 
$$E\left(\frac{d\til{Q}}{dP}\bigg|\scr{F}_t\right)=\frac{1}{\alpha} u_t'(c_t).$$ 
\end{bem}

\section*{Acknowledgements}
We thank two anonymous referees for careful reading of the manuscript and numerous constructive comments.

\end{document}